\documentstyle[preprint,floats,prd,aps]{revtex}
\input epsf.tex
\def\DESepsf(#1 width #2){\epsfxsize=#2 \epsfbox{#1}}
\begin{document}
\preprint{\vbox{\hbox{OCHA-PP-64}}  }
\draft
\title{Examination of the Resonance Contributions to Dileptonic Rare
B-Decays
\footnote{Work supported by the Japanese Society for the Promotion of
Science.}} %
\author{Mohammad R. Ahmady}\address{Department of Physics \\
Ochanomizu University \\
1-1 Otsuka 2, Bunkyo-ku,Tokyo 112, Japan}

\date{July 1995}
\maketitle
\begin{abstract}
We analyse the long-distance contribution to $B\to X_s\ell^+\ell^-$
differential decay rate when the momentum dependence of
$\psi$ and $\psi'$-$\gamma$ conversion strength is taken into account.
The results indicate that the resonance to nonresonance interference in
the dilepton invariant mass distribution is substantially reduced.
\end{abstract}
Rare B-decays have been the focus of extensive experimantal and
theoretical investigations.  The observation of the inclusive $b\to
s\gamma$ decay and the exclusive mode $B\to K^*\gamma$ by CLEO
collaboration
\cite{cleo} has risen the hope that other flavor changing neutral current
(FCNC) transitions, like $b\to s\ell^+\ell^-$ $(\ell =e\; ,\; \mu )$ for
instance, will be within experimental reach in near future.  The
measurement of these processes can serve, among other things, to shed
light on some less-known CKM matrix elements, as well as, imposing
constraints on ''new Physics'' beyond the Standard Model(SM).  However,
in order to make any reliable conclusion on these from the experimantal
measurements of rare B-decays, we must improve our understanding of
the theoretical uncertainties in the calculation of these processes.

One of the main background sources for rare B-decays is the
long-distance(LD) contributions to these processes. A lot of theoretical
attention has been focused on this subject \cite {ld,lms,dtp,dht,eims,osht},
due to
the fact that without a reliable estimate of the LD contributions one can
not draw accurate conclusions from experimental results.

The CKM favored resonance contributions to $b\to s$ transition is due to the
conversion of the intermediate $\psi (NS)$ vector mesons to real($b\to
s\gamma$ decay) or virtual($b\to s\ell^+\ell^-$ decay) photons.  The
momentum dependence of $\psi -\gamma$ conversion strength was
investigated long time ago in Ref \cite{terasaki} in order to explain the
data on $\psi$ leptonic width and photoproduction simultaneously.
Recently, this has been pointed out again in Refs \cite{dht,eims}, where a
large suppression of $\psi -\gamma$ transition on photon mass shell, is
argued to indicate that the LD contribution to $b\to s\gamma$ decay could
be substantially smaller than previous estimates.  For $b\to
s\ell^+\ell^-$ decay however, the momentum dependence of $\psi
-\gamma$ conversion strength has not been taken into account up to now.

In this paper, we analyze the resonance contribution to the inclusive
dileptonic
rare B-decays using a momentum dependent $\psi -\gamma$ conversion
strength.  We show that the dileptonic mass distribution is indeed
sensitive to the short-distance(SD) contributions for a broader
$q^2$($q^2$ is the invariant dileptonic mass).

We start with the low energy effective Lagrangian for $b\to
s\ell^+\ell^-$ \cite{amm,gsw}
\begin{equation}
\displaystyle
L_{eff} =\frac {G_F}{\sqrt 2} \left ( \frac {\alpha }{4 \pi s_W^2}
\right ) V^*_{ts}V_{tb}
(A\bar s L_\mu b \bar \ell L^\mu \ell +B \bar s L_\mu b \bar\ell
R^\mu \ell +2m_b s_W^2F \bar s T_\mu b \bar\ell \gamma^\mu\ell ),
\end{equation}
where
$$
L_\mu =\gamma_\mu (1-\gamma_5), \quad
R_\mu =\gamma_\mu (1+\gamma_5),
$$
and
$$
T_\mu = -i\sigma_{\mu \nu} (1+\gamma_5) q^\nu /q^2 .
$$
$V_{ij}$ are the Cabibbo-Kobayashi-Maskawa matrix elements,
$s_W^2=sin^2\theta_W\approx 0.23$ ($\theta_W$ is the weak angle), $G_F$
is the Fermi constant and $q$ is the total momentum of the final
$\ell^+\ell^-$ pair.

The SD parts of $A$ and $B$, denoted by $A^{\rm SD}$ and $B^{\rm SD}$, arise
from W box diagrams and penguin diagrams with Z gauge boson and photon
coupled to $\ell^+\ell^-$ pair.  For these coefficients and $C=s_W^2F$ we
have:
\begin{equation}
\begin{array}{rl}
A^{\rm SD}=& {\bar C}^{\rm Box}(x_t)+{\bar C}^Z(x_t)+B^{\rm SD} \\
\displaystyle
B^{\rm SD} =& -s_W^2\left [F_1^{(s)}(x_t)+2{\bar C}^Z(x_t)-\frac{4}{9}(\ln
x_t +1) \right .\\ &
\displaystyle \left .
+\frac{4\pi}{\alpha_s(M_W)}\left \{
-\frac{4}{33}(1-\eta^{-11/23})+\frac{8}{87}(1-\eta^{-29/23})\right
\}C_2(M_W)\right ] \\
C =&
\displaystyle -s_W^2\left \{ \eta^{-16/23}\left [
F_2(x_t)-\frac{116}{135}(\eta^{10/23}-1)C_2(M_W)-\frac{58}{189}(\eta^{28/23}-1)C_2
(M_W)\right ]\right \}

\end{array}
\end{equation}
Where $x_t=m_t^2/M_W^2$, $\eta =\alpha_s(m_b)/\alpha_s(M_W)$ and \cite
{il}
\begin{equation}
\begin{array}{rl}
{\bar C}^Z(x)=&
\displaystyle
\frac{1}{4}x+\frac{3}{8}\frac{x}{1-x}+\frac{3}{8}\frac{2x^2-x}{{(1-x)}^2}\ln
(x) \\
{\bar C}^{\rm Box}(x)+{\bar C}^Z(x)=&
\displaystyle
\frac{1}{4}x+\frac{3}{4}\frac{x}{1-x}+\frac{3}{4}{\left (\frac{x}{1-x}\right
)}^2\ln (x)  \\
F_1^{(s)}(x) =&
\displaystyle
\frac{63x-151x^2+82x^3}{36{(1-x)}^3}+\frac{63x-138x^2+59x^3+10x^4}{36{(1-x)}^4}
\ln (x)\\
F_2(x) =&
\displaystyle
\frac{7x-5x^2-8x^3}{12{(1-x)}^3}+\frac{2x^2-3x^3}{2{(1-x)}^4}\ln (x)
\end{array}
\end{equation}
$C_2$ is the Wilson coefficient of the four fermion operator with
$C_2(M_W)=-1$.  For our numerical evaluation we use $m_t=180$ GeV (which is the
weighted average of the recent CDF and D0 results \cite{top}) and
$\Lambda_{QCD}=100$ MeV so that $\alpha_s(M_W)=0.12$ and $\eta =1.75$ (for
$m_b=4.5$ GeV).  As a result, we obtain:
\begin{equation}
\begin{array}{l}
A^{\rm SD}=2.020 ,\\
B^{\rm SD}=-0.173 ,\\
C=-0.146 .
\end{array}
\end{equation}

The LD contributions enter $A$ and $B$ coefficients through charm quark loop
($c\bar c$ continuum), and the resonance
contributions from $\psi$ and $\psi'$.
\begin{equation}
A^{\rm LD}=B^{\rm LD}=-s_W^2\left (3C_1(m_b)+C_2(m_b)\right )(\tau^{\rm
cont}+\tau^{\rm res})
\end{equation}
$C_1(m_b)$ and $C_2(m_b)$ are the QCD corrected Wilson
coefficients:
\begin{equation}
\begin{array}{rl}
C_1(m_b)=&
\displaystyle
\frac{1}{2}\left (\eta^{-6/23}-\eta^{12/23}\right )C_2(M_W) \\
C_2(m_b)=&
\displaystyle
\frac{1}{2}\left (\eta^{-6/23}+\eta^{12/23}\right )C_2(M_W) \\
\end{array}
\end{equation}
The $c\bar
c$ continuum contribution is obtained from the electromagnetic penguin
diagrams \cite {gsw} \begin{equation}
\displaystyle
\tau^{\rm cont}=g\left
(\frac{m_c}{m_b}\; ,\; z\right )
\end{equation}
where $z=q^2/m_b^2$ and
\begin{equation}
\displaystyle
g(y\; ,\; z)= \left \{ \begin{array}{ll}
\displaystyle
-\left [ \frac{4}{9}\ln
(y^2)-\frac{8}{27}-\frac{16}{9}\frac{y^2}{z}+\frac{2}{9}\sqrt{1-\frac{4y^2}{z}}
\left (2+\frac{4y^2}{z}\right )\left (\ln \frac{
\vert 1+\sqrt{1-\frac{4y^2}{z}}\vert}{\vert 1-\sqrt{1-\frac{4y^2}{z}}\vert}
+i\pi\right )\right ] & \;  z\ge 4y^2 \\
\displaystyle
-\left [ \frac{4}{9}\ln
(y^2)-\frac{8}{27}-\frac{16}{9}\frac{y^2}{z}+\frac{4}{9}\sqrt{\frac{4y^2}{z}-1}
\left (2+\frac{4y^2}{z}\right ) arctan
\frac{1}{\sqrt{\frac{4y^2}{z}-1}}\right ]  & \;  z\le 4y^2

\end{array}
\right.
\end{equation}

On the other hand, the resonance contributions from $\psi$ and $\psi'$
can be incorporated by using a Breit-Wigner form for the resonance
propagator \cite{lms,dtp}:
\begin{equation}
\tau^{\rm res}=\frac{16\pi^2}{9}\left (\frac{f^2_\psi
(q^2)/m^2_\psi}{m^2_\psi-q^2-im_\psi\Gamma_\psi}+\; (\psi\to\psi')\right
)e^{i\phi}
\end{equation}
The relative phase $\phi$ that determines the sign between $\tau^{\rm
cont}$ and $\tau^{\rm res}$ is chosen to be zero due to unitarity
constraint \cite {ot}.

The Wilson coefficient sum $3C_1(\mu )+C_2(\mu )$ is very sensitive to
QCD scale parameter $\Lambda_{QCD}$ as well as the renormalization point
$\mu$ \cite{wilco}.  Inserting $\eta =1.75$, one would obtain
$3C_1(m_b)+C_2(m_b)=-0.389$.  However, if we treat this combination of
the Wilson coefficients as a phenomenological parameter, a value
$$
\vert 3C_1(m_b)+C_2(m_b)\vert =0.72
$$
fits the data on the semi-inclusive $B\to X_s\psi$ \cite{dht}.  We use
this phenomenological value throughout our calculations.

On the other hand, from (9) we observe that $\tau^{\rm res}$ depends
quadratically on $f_V(q^2) (V=\psi\;,\;\psi' \;)$ defined as:
\begin{equation}
<0|\bar c\gamma_\mu c|V (q)>=f_V(q^2)\epsilon_\mu
\end{equation}
where $\epsilon_\mu$ is the polarization vector of the vector meson $V$.
As we mentioned
earlier, it has been pointed out recently that in the context of Vector
Meson Dominance, data on photoproduction of $\psi$ indicates a large
suppression of $f_\psi (0)$ compare to $f_\psi (m^2_\psi )$ \cite {dht}.
This has been confirmed independently in Ref \cite {eims} by constraining
the dominant LD contribution to $s\to d\gamma$ using the present upper
bound on the $\Omega^-\to\Xi^-\gamma$ decay rate.  In fact, it is argued
that this large suppression results in a much smaller LD contribution
to $b\to s\gamma$ transition.

In the dileptonic rare B-decays, however, the momentum dependence of
$f_V$(or equivalently, $\psi -\gamma$ transition) has not
been taken into account up to now, and $f_V(q^2)$ is normally replaced with
the decay constant $f_V(m^2_V )$ obtained from the leptonic width of
$\psi$ and $\psi'$:
$$
\Gamma (V\to\ell^+\ell^- )=\frac{16\pi\alpha^2}{27m_V^3}f_V^2(m^2_V)
$$
The spectrum obtained this way, is dominated by the resonance
interference for a broad range of $q^2$, as already noted in the
literature \cite {osht,amm}.

In this work,however, we use a momentum dependent $f_V(q^2)
(V=\psi\; ,\; \psi' )$ in $\tau^{\rm res}$ (as in Ref \cite {dht}, we assume
that the same suppression occurs for $\psi'$). Of course, there is no
significant change in the total branching ratio which is dominated by
$\psi$ and $\psi'$ resonance contributions which in turn is due to the
fact that the dileptonic mass spectrum has peaks at $q^2=m_\psi^2\; ,\;
m^2_{\psi'}$.  However, as we demonstrate later on, as a result, the
resonance to
nonresonance interference is substantially reduced, leaving a
broader region of invariant
mass spectrum sensitive to a large extent to SD physics.

We use the momentum dependent $f_V(q^2)$($V=\psi\; ,\;\psi'$) derived in Ref
\cite {terasaki} based on the intermediate quark and antiquark state:
\begin{equation}
f_V(q^2)=f_V(0)\left (1+\frac{q^2}{c_V}\left [d_V -h(q^2)\right ]\right )
\end{equation}
where $c_\psi =0.54\; ,\; c_{\psi'}=0.77$ and $d_\psi =d_{\psi'}=0.043$.
$h(q^2)$ is obtained from a dispersion relation involving the imaginary
part of the quark-loop diagram:
\begin{equation}
\displaystyle
h(q^2)=\frac{1}{16\pi^2r}\left \{
-4-\frac{20r}{3}+4(1+2r)\sqrt{1-\frac{1}{r}}arctan\frac{1}{\sqrt{1-\frac{1}{r}}}\right \}
\end{equation}
with $r=q^2/4m_q^2$ for $0\leq q^2\leq 4m_q^2$.  $m_q$ is the effective
quark mass and assuming that the vector mesons are weakly bound systems of a
quark and an antiquark, we take $m_q\approx m_V/2$.  As a result, eqn (11),
defined for $0\leq q^2\leq m_V^2$, is an interpolation of $f_V$ from the
experimental data on $f_V(0)$(from photoproduction) and $f_V(m_V^2)$(from
leptonic width) based on quark-loop
diagram.  We assume $f_V(q^2)=f_V(m_V^2)$ for $q^2 > m_V^2$ mainly due
to the fact that the behavior of $\psi -\gamma$ conversion strength is
not clear in this region.  In any case, our focus will be on the
invariant mass spectrum region bellow $m_\psi^2$ where the effect of
momentum-dependent $\psi -\gamma$ transition is more significant.

The differential decay rate for $b\to X_s\ell^+\ell^-$, taken as
the free quark decay $b\to s\ell^+\ell^-$, can be written as:
\begin{equation}
\begin{array}{rl}
\displaystyle\frac {1}{\Gamma (B\to X_c e\bar\nu )}\frac {d\Gamma}{dz} (B\to
X_s\ell^+\ell^- ) =& \displaystyle
\left ( \frac {\alpha}{4\pi s_W^2} \right )^2
\frac {2}{f(m_c/m_b)}
\frac {{\vert V_{ts}^*V_{tb}\vert}^2}{{\vert V_{cb}\vert}^2}
{(1-z)}^2 \\ \times &
\displaystyle
\left       ( ( \vert  A \vert^2 +{\vert  B \vert}^2)(1+2z)
                +2\vert  C \vert^2 (1+2/z)\right. \\
     \; &
\displaystyle
\left. +  6Re[{( A + B)}^* C]\right ) \\
\end{array}
\end{equation}
where
$$
f(x)=1-8x^2+8x^6-x^8-24x^4ln(x).
$$
By normalizing to the semileptonic rate in (13), the strong
dependence on the b-quark mass cancels out.  In fig.1, we show the
invariant dilepton mass spectrum corresponding to eqn(13) for cases when
i) the resonance term $\tau^{\rm res}$ is not included, ii) $\tau^{\rm res}$
is included but $f_V(q^2)$ in eqn (9) is replaced with constant
$f_V(m_V^2)$ and finally, iii) $\tau^{\rm res}$ with $f_V(q^2)$ inserted
from eqn (11) is included.  From fig.1 we observe  that the resonance to
non-resonance interference in the invariant mass spectrum, which is
measured by the deviation from the nonresonant spectrum(thin line), is
suppressed
considerably due to momentum-dependence of $\psi -\gamma$ conversion
strength.  For example, at $q^2/m_b^2\approx 0.3$, the resonance
interference amounts to around $2\%$ of the differential branching ratio
as compared to $20\%$ in the case where fixed $f_V(m_V^2)$ is used.  As a
result, we believe, contrary to previous conclusions \cite {osht},
dilepton mass spectrum can be used for testing SD physics without
significant resonance interference.

\vskip 2.0cm
{\bf \Large Acknowledgement}
\vskip 0.5cm
The author would like to thank R. R. Mendel for useful discussions.

\newpage
\begin{flushleft}
\bf \Huge Figure Caption
\end{flushleft}
{\bf \Large Figure 1:}
The dileptonic invariant mass spectrum for the decay
$b\to
s\ell^+\ell^-$.  The thin, dotted and bold lines correspond to spectrum
without resonances, with resonances but constant $V-\gamma$ conversion
strength and with resonances having momentum dependent $V-\gamma$
transition respectively.

 \newpage
\setlength{\unitlength}{0.240900pt}
\ifx\plotpoint\undefined\newsavebox{\plotpoint}\fi
\sbox{\plotpoint}{\rule[-0.200pt]{0.400pt}{0.400pt}}%


 
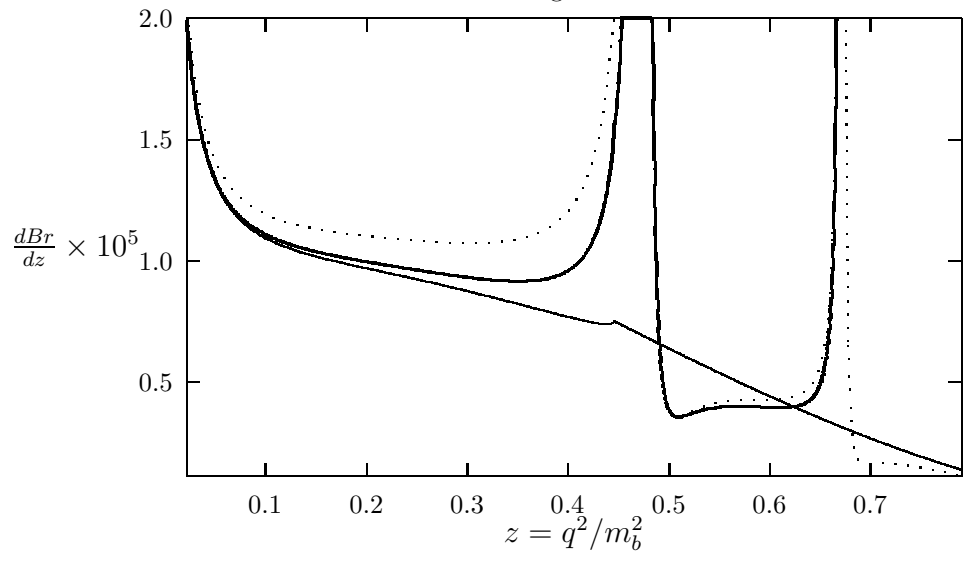
\end{document}